\documentclass[a4paper,12pt]{article}
\usepackage[pctex32]{graphics}

\begin{document}
\topmargin 0pt
\oddsidemargin 0mm

\newcommand{\alp}{\alpha}
\newcommand{\bta}{\beta}
\newcommand{\gmm}{\gamma}
\newcommand{\del}{\delta}
\newcommand{\omg}{\omega}
\newcommand{\sgm}{\sigma}
\newcommand{\lmd}{\lambda}
\newcommand{\tha}{\theta}
\newcommand{\vph}{\varphi}
\newcommand{\Alp}{\Alpha}
\newcommand{\Bta}{\Beta}
\newcommand{\Gmm}{\Gamma}
\newcommand{\Del}{\Delta}
\newcommand{\Omg}{\Omega}
\newcommand{\Sgm}{\Sigma}
\newcommand{\Lmd}{\Lambda}
\newcommand{\Tha}{\Theta}
\newcommand{\half}{\frac{1}{2}}
\newcommand{\rnd}{\partial}
\newcommand{\nab}{\nabla}

\newcommand{\beqn}{\begin{eqnarray}}
\newcommand{\eeqn}{\end{eqnarray}}
\newcommand{\be}{\begin{equation}}
\newcommand{\ee}{\end{equation}}

\begin{titlepage}

\vspace{5mm}
\begin{center}
{\Large \bf Entropic force versus  temperature force } \vspace{12mm}

{\large   Yun Soo Myung \footnote{e-mail
 address: ysmyung@inje.ac.kr}, Hyung Won Lee\footnote{e-mail address hwlee@inje.ac.kr}, and  Yong-Wan Kim\footnote{e-mail
 address: ywkim65@gmail.com}}
 \\
\vspace{10mm} {\em  Institute of Basic Science and School of
Computer Aided Science, Inje University, Gimhae 621-749, Republic of
Korea}

\end{center}

\vspace{5mm} \centerline{{\bf{Abstract}}}
 \vspace{5mm}
We introduce the  cavity enclosing  a source mass $M$ to define the
temperature force. Starting with the Tolman temperature in the
stationary spacetime, we find a non-relativistic temperature
$T_{non}= T_\infty(1-\Phi/c^2)$ with the Newtonian potential $\Phi$.
This temperature could be also derived from the Tolman-Ehrenfest
effect, satisfying a relation of $T=T_{\infty}e^{-\Phi/c^2}$ with
the local temperature $T$. Finally, we derive the temperature force
$\vec{F}_{tem}=mc^2(\vec{\nabla} \ln T )$ which leads to the
Newtonian force law without introducing the holographic screen
defined by holographic principle and equipartition law for entropic
force.

\end{titlepage}
\newpage
\renewcommand{\thefootnote}{\arabic{footnote}}
\setcounter{footnote}{0} \setcounter{page}{2}

\section{Introduction}
Since the discovery of the laws of black hole
thermodynamics~\cite{BCH}, Bekenstein~\cite{Bek} and
Hawking~\cite{Hawk} have suggested a deep connection between gravity
and thermodynamics, realizing black hole entropy and Hawking
radiation.  Later on, Jacobson~\cite{Jac} has demonstrated that
Einstein equations of  general relativity (describing relativistic
gravitation) could be derived by combining general thermodynamic
pictures with the equivalence principle. Padmanabhan~\cite{Pad3} has
observed that the equipartition law for  horizon degrees of freedom
combined with the Smarr formula leads to the Newton's law of
gravity.  This may imply  that the entropy is to link general
relativity
 with the statistical description of unknown spacetime microscopic
 structure when the horizon is  present.

Recently, Verlinde has proposed the Newtonian force law as an
entropic force (non-relativistic version)  by using  the holographic
principle and  the equipartition rule in the absence of
horizons~\cite{Ver}.
 If it
is proven correct, gravity is not a fundamental interaction, but an
emergent phenomenon which arises from the statistical behavior of
microscopic degrees of freedom encoded on a holographic screen. In
other words, the force of gravity is not something ingrained in
matter itself, but it is an extra physical effect, emerging from the
interplay of mass, time and space (information) through the entropy.

However, an urgent question  is {\it how one can construct a
spherically  holographic screen of radius $r$ which encloses a
source mass $M$ located at the origin using the holographic
principle.} This is a very important issue~\cite{MK} because the
holographic screen (an exotic description of spacetime) originates
from relativistic approaches to black hole ~\cite{Hoo,Suss} and
cosmology~\cite{Bou,SL}. Verlinde has introduced this screen  by
analogy with an absorbing process of a particle around the event
horizon of black hole. Considering a smaller test mass $m$ located
at $\Delta x$ away from the screen and getting the change of entropy
on the  screen,  its behavior should resemble that of a particle
approaching a stretched horizon of a black hole, as was described by
Bekenstein~\cite{Bek}. It is clear that  Verlinde has introduced the
holographic screen as a basic input to derive the entropic force.

The next important question is {\it why the equipartition rule could
be applied to this non-relativistic  screen to define the
temperature  without any justifications.}  For black holes, the
equipartition rule becomes the Smarr formula of $E=NT/2=2ST$ when
using $N=4S=\frac{Ac^3}{G \hbar}$. Also it can be derived from the
first law of thermodynamics $dE=TdS$ for the Schwarzschild black
hole where the Komar charge is just the ADM mass $M$.  Even though
the equipartition rule is available for the classical
thermodynamics, the holographic principle of $N=Ac^3/G\hbar$ is not
guaranteed to apply to any non-relativistic situations.  In this
sense, this issue is  closely related to the first one.

If the above two questions are answered properly, one would  make a
further step to understand the origin of Newtonian  force.  However,
there exists still a gap between non-relativistic  approach (absence
of horizons) and relativistic approach (presence of horizons). It is
not legitimate to use the holographic screen defined by using the
holographic principle and the equipartition rule in deriving a
non-relativistic force law.
 It is
shown that Verlinde  has used  some ideas for obtaining Einstein
equations due to Jacobson's derivation of Einstein equations. Also,
it seems  that he was using circular reasoning in his equations, by
starting out with Einstein gravity.

In this work, we will show  that the Newtonian force law could be
derived from the temperature force on the cavity without referring
to holographic principle  and equipartition rule. This
 is a natural approach because it starts with the Tolman
temperature on the stationary (black hole) spacetime and it is
closely related to the Tolman-Ehrenfest effect which states that
{\it even for the Newtonian gravity, the temperature is not constant
in space at equilibrium}~\cite{Tol1,Tol2}.

\section{Entropic force}

In this section, we briefly review how the Newtonian  force law
emerges from  entropic considerations~\cite{Ver}.  Explicitly, when
a test particle with mass $m$ is located near a holographic screen
with distance $\Delta x$, the change of entropy on a holographic
screen may take the form \be \label{eq1} \Delta S= 2\pi k_B  \Delta
x \frac{m c}{\hbar}. \ee Considering that the entropy of a system
depends on the distance $\Delta x$, an entropic force $F_{ent}$
could be arisen from the thermodynamical conjugate of the distance
as \be \label{eq2} F_{ent} \Delta x=T_{hs} \Delta S \ee which may be
regarded as an indication that the first law of thermodynamics is
realized  on the holographic screen.  Plugging (\ref{eq1}) into
(\ref{eq2}) leads to an important connection between the entropic
force and temperature on the screen \be \label{eq3}
F_{ent}=\frac{2\pi k_B m c}{\hbar} T_{hs}. \ee One uses mainly this
connection  to derive the entropic force, only after setting the
temperature $T_{hs}$ on the holographic screen.

Introducing the Unruh temperature~\cite{Unruh} as the holographic
screen temperature \be T_{hs} \to T_U=\frac{\hbar}{2\pi k_B c} a,
\ee one may find the second law \be F_{ent}=ma. \ee However, the
Unruh temperature has originated at the relativistic quantum field
theory. It seems that the Unruh effect remains a theoretical
phenomena with some restrictions~\cite{KP} and this effect is not
available for a non-relativistic case. Therefore, it is problematic
to use the Unruh temperature as the screen temperature.

In order to derive  the holographic screen temperature $T_{hs}$, let
us suggest that the energy $E$ is distributed on a spherical screen
with radius $r$ and the source mass $M$ is located at the origin of
coordinates. Then, we assume that the equipartition
rule~\cite{Pad2,Pad3}, the equality of energy and mass, and the
holographic principle, respectively, hold as \be \label{eq4}
E=\frac{1}{2 }N k_B T_{hs},~~E=Mc^2,~~N=\frac{Ac^3}{G\hbar}=4S \ee
with the area of a holographic screen $A=4\pi r^2$. Importantly,
these should be combined to put   the temperature on the screen \be
\label{eq5} T_{hs}=\frac{\hbar}{2\pi k_B c}\frac{GM}{r^2}.\ee
Substituting (\ref{eq5}) into (\ref{eq3}), one obtains the Newtonian
force law as the entropic force\be \label{eq6} F_{ent}=\frac{G m
M}{r^2}=mg, \ee where the Galilean acceleration of the Newtonian
gravity is given by \be g=\frac{GM}{r^2}=\frac{2\pi k_B c}{\hbar}
T_{hs}. \ee

We summarize ``how the entropic force is realized": as a test
particle with mass $m$ approaches the holographic screen, its own
entropy bits begin to transfer to the holographic screen, and it is
this increase in the screen entropy that generates an attractive
force on the test mass.  We must  emphasize that as is shown in
(\ref{eq3}), the role of holographic screen temperature  $T_{hs}$ is
essential for deriving the entropic force. Unless the Newtonian
force is emergent as the entropic force, the choice of holographic
screen temperature seems to be a major flaw of Verlinde's idea.
Instead, we use the Tolman-Ehrenfest effect which states that  for
the Newtonian gravity, the temperature is not constant in space at
equilibrium. Thus, we could define the local temperature on any
place where the test particle is located.

\section{Temperature force on the  cavity }
In the presence of horizons, it is natural to define the horizon
temperature as the surface gravity of event horizon for black hole
and apparent horizon for cosmology. In the absence of horizons,
Verlinde has attempted to introduce the holographic screen
temperature instead of horizon temperature~\cite{Ver}.  However,  as
was emphasized previously, the usage of the holographic screen is
not guaranteed to describe  a non-relativistic case of a source mass
$M$. Hence, if the Newtonian gravity is not a fundamental force, it
would be better to describe  the Newtonian force law without
imposing the holographic principle and equipartition rule.

In this section, we wish to define a temperature on the cavity
enclosing the source mass $M$ located at the origin without imposing
the holographic screen.  First we start with  a stationary
spherically symmetric spacetime which may include a black hole \be
\label{eq7}
ds^2_{stat}=g_{00}(\vec{x})dx^0dx^0-g_{ij}(\vec{x})dx^idx^j. \ee
 The
Tolman temperature observed by an observer located on the cavity is
defined by~\cite{TolmanT} \be \label{eq8} T_T(r)=
\frac{T_{\infty}}{\sqrt{g_{00}}},~~g_{00}(r)=1+\frac{2\Phi(r)}{c^2}
\ee where $ T_\infty$ is the temperature  measured by observer at
infinity and the denominator of $\sqrt{g_{00}}$ is considered as the
redshift factor. $\Phi$ is the Newtonian potential. The situation
becomes clear when introducing a Schwarzschild black
hole~\cite{York}: \be
g_{00}=1-\frac{2MG}{c^2r}=1+\frac{2\Phi}{c^2},~~T_\infty=T_H=\frac{1}{8\pi
GM} \ee where $T_H$ is the Hawking temperature observed at the
infinity. It is worth to mention  a non-relativistic  case of $c \to
\infty$ for Newtonian gravity \be T_{T}\simeq
\frac{T_\infty}{1+\frac{\Phi}{c^2}}\simeq
T_\infty\Big(1-\frac{\Phi}{c^2}\Big)\equiv T_{non}, \ee where the
last relation defines the non-relativistic temperature $T_{non}$. On
the other hand, from the Tolman-Ehrenfest effect~\cite{Tol2},
equilibrium between two systems happens when the total entropy is
maximized as \be dS=dS_1+dS_2=0. \ee  If a heat element $dE_1$
enters the second system, then it leads to the condition of thermal
equilibrium \be \label{there}\frac{dS_1}{dE_1}-\frac{dS_2}{dE_2}=0
\to T_1=T_2. \ee
\begin{figure}[t!]
\centering
\includegraphics{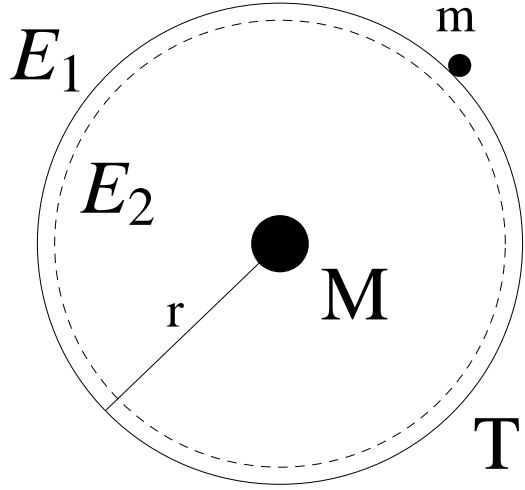}
\caption{Cavity with temperature $T$ is placed at a distance of $r$
from the source mass $M$ at the origin. A test mass with $m$ is on
the cavity. Solid and dotted cavities are introduced to realize the
Tolman-Ehrenfest effect such as  Eq.(\ref{e2e1}).} \label{fig1.eps}
\end{figure}
 However, if
two systems are at different Newtonian potentials, the amount of
energy $dE_1$ leaving the first one does not equal the amount of
energy entering the second one. This is because $E=mc^2$ and the
equality of inertial and gravitational mass  imply that any form of
energy has a gravitational mass and thus, falls in a gravitational
field. In this case, as is shown in Fig. 1, it is very  natural to
define the local temperature $T(r)$  on  the cavity  located at $r$
from the source mass $M$.    $dE_2$ is $dE_1$ increased by the
potential energy $m \Delta \Phi$: \be \label{e2e1} dE_2=dE_1
\Big(1+\frac{\Delta \Phi}{c^2} \Big). \ee Then, making use of
(\ref{there}) and (\ref{e2e1}),  one finds \be
\frac{1}{T_2}=\frac{dS_2}{dE_2}=-\frac{dS_1}{dE_2}=\frac{1}{T_1}\frac{1}{1+\frac{\Delta
\Phi}{c^2}}\ee which implies\footnote{As was explained in
Ref.\cite{RS}, $\vec{g}$ is the Galilean acceleration of gravity and
$1/c^2$ is inserted as a slight relativistic  effect.
$\vec{\nabla}\ln T=\vec{g}/c^2$ means that a vertical column of
fluid at equilibrium is hotter at the bottom. For example,
$\vec{\nabla}\ln T=10^{-18}/cm$ on the surface of the Earth with
$g=9.81 m/s^2$. We should note that even though the derivation of
$\vec{\nabla}\ln T=\vec{g}/c^2$ is different from here, it has
already been derived in Ref.\cite{Tol1}. } \be \frac{\Delta
\Phi}{c^2} =\frac{T_2}{T_1}-1 \to \frac{\vec{\nabla}
\Phi}{c^2}=-\frac{\vec{\nabla}T}{T}=-\frac{\vec{g}}{c^2}. \ee Here
we have made replacements: $T_1 \to T$ and $T_2-T_1 \to - \Delta T$.
Then, the above leads to an important relation between temperature
and Newtonian potential \be -\Phi= c^2 \ln [T/T_\infty] \ee which
provides the local temperature as a function of Newtonian potential
\be T(r)=T_\infty e^{-\Phi/c^2}\simeq T_\infty\Big(
1-\frac{\Phi}{c^2}\Big). \ee As is depicted in Fig. 2, the
non-relativistic temperature $T_{non}$  is the first order
approximation of the local temperature $T(r)$. Two are different at
small $r$, while two are nearly the same for large $r$ and approach
$T_\infty$ at infinity. The connection to the Newtonian potential is
shown explicitly.

\begin{figure}[t!]
\centering
\includegraphics{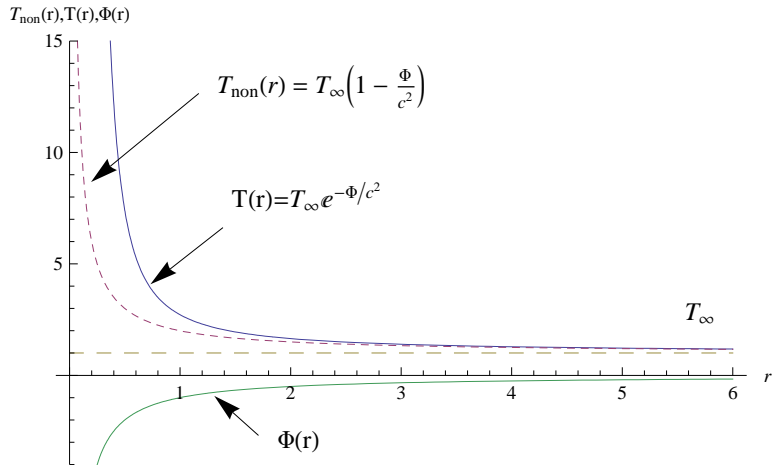}
\caption{Non-relativistic temperature $T_{non}(r)$(dotted curve)
 and local temperature $T(r)$ (solid curve)
are depicted in the upper plane.  $T_\infty$ is the undetermined
temperature observed  by the distant observer within the Newtonian
gravity. The Newtonian potential $\Phi(r)$ is drawn as function of
$r$ in the lower plane.} \label{fig2.eps}
\end{figure}

Finally, the temperature force is defined as \be \vec{F}_{tem}=mc^2
\Bigg(\frac{\vec{\nabla}T}{T}\Bigg)=mc^2 \vec{\nabla} \ln T= m
\vec{g}, \ee which leads to the Newtonian universal law. We wish to
explain how the temperature force comes into play. In contrast to
the entropy change in the holographic screen when a test mass
approaches the screen, the energy change is essential in the cavity
when a test approaches the cavity.   This is because $E=mc^2$ and
the equivalence principle (equality of inertial and gravitational
mass) imply  that any form of energy has a gravitational mass and
thus, falls in a gravitational field.  Hence, $dE_2$ is $dE_1$
increased by the potential energy $m \Delta \Phi$, which generates
an attractive force on the test mass. The key mechanism is
different: the entropic force is realized as \be {\rm test~
mass}~(m)~ \to {\rm entropy~ increase~ in~ holographic ~screen},\ee
while the temperature force is realized as \be {\rm test~mass}~(m)
\to {\rm energy~ increase ~in ~cavity}.\ee
 Also, the entropic force
depends on the screen temperature itself, while the temperature
force depends on the gradient of logarithmic  temperature. The
Newtonian potential is realized by the entropy in the entropic
force, whereas it is realized by the temperature in the temperature
force. In this sense, we use the notion of  temperature force
instead of the entropic force. In defining the temperature force,
$T_\infty$ is not uniquely determined, while $\hbar$ and $k_B$ do
not appear, compared to the entropic force.

\section{Discussions}
\begin{table}
 \caption{Comparison between entropic force (EF) and temperature force (TF). HP(ER)
 denote holographic principle (Equipartition rule), and T-EE means Tolman-Ehrenfest effect. Finally, HS denotes holographic screen.}

\begin{tabular}{|c|c|c|c|c|c|}
  \hline
   &principle& where & temperature & potential & force \\
  \hline
 EF & HP and ER&  HS & $T=\frac{\hbar}{2\pi k_B c} g$ & $\Phi \propto S $& $F=\frac{2\pi k_B m c}{\hbar}T=mg$ \\ \hline
 TF &T-EE& cavity & $\vec{\nabla} \ln T=\frac{\vec{g}}{c^2}$  & $\Phi \propto \ln T$ &$\vec{F}=mc^2\vec{\nabla}\ln T=m\vec{g}$\\ \hline
  \hline
\end{tabular}
\end{table}
It is fare to say that the origin of the gravity is  not yet fully
understood. If the gravity is not a fundamental force, it may be
emergent from the other approach to gravity. Verlinde's idea was
that Newtonian force law could be emergent from the equipartition
rule and the holographic principle~\cite{Ver}.  However, an
important thing  is to show that  the holographic screen could be
defined by enclosing a source mass $M$.  This is unlikely possible
to occur.

 In this work, we have defined the
local temperature on the  cavity from the Tolman-Ehrenfest effect.
It is a  natural way to define the temperature without imposing the
equipartition rule and the holographic principle. We have introduced
the  cavity enclosing  a source mass $M$ to define the temperature
force. Starting the Tolman temperature
$T_{T}=T_\infty/\sqrt{g_{00}}$, we find a non-relativistic
temperature $T_{non}= T_\infty(1-\Phi/c^2)$. This temperature could
be also derived from the Tolman-Ehrenfest effect, satisfying a
relation $T=T_{\infty}e^{-\Phi/c^2} \simeq T_\infty(1-\Phi/c^2)$.
Finally, from the defining equation $\vec{F}_{tem}=mc^2(\vec{\nabla}
\ln T)$, we derive the temperature force which is identified with
the Newtonian force law.

As is shown in Table 1, there are two kinds of forces which led to
the same Newtonian universal law. One is based on the entropy of
holographic screen, while the other is based on the local
temperature on the cavity.   Unless the Newtonian force is emergent
as the entropic force, the choice of holographic screen temperature
seems to be a major flaw of Verlinde's idea. On the other hand, our
idea on the temperature force  came from the Tolman temperature
which is  a well-defined quantity in the relativistic description.
Hence, we did not need to introduce an exotic spacetime of the
holographic screen defined by holographic principle and
equipartition rule.

\section*{Acknowledgment}
 This work  was supported by Basic Science Research
Program through the National  Research  Foundation (NRF) of  Korea
funded by the Ministry of Education, Science and Technology
(2009-0086861).

\end{document}